\documentclass[times, 10pt,twocolumn]{article}%
\usepackage{latex8}
\usepackage{amsmath}
\usepackage{amsfonts}
\usepackage{amssymb}
\usepackage{graphicx}
\usepackage{times}%
\newtheorem{theorem}{Theorem}

\newtheorem{definition}[theorem]{Definition}

\newtheorem{lemma}[theorem]{Lemma}

\newtheorem{proposition}[theorem]{Proposition}

\newenvironment{proof}[1][Proof]{\noindent\textbf{#1.} }{\ \rule{0.5em}{0.5em}}
\begin{document}

\title{$\mathsf{QMA/qpoly}\subseteq\mathsf{PSPACE/poly}$: De-Merlinizing Quantum Protocols}
\author{Scott Aaronson\thanks{Email: scott@scottaaronson.com. \ Supported by ARDA,
CIAR, and IQC. \ Part of this work was done at Caltech.}\\University of Waterloo}
\date{}
\maketitle

\begin{abstract}
This paper introduces a new technique for removing existential quantifiers
over quantum states. \ Using this technique, we show that there is no way to
pack an exponential number of bits into a polynomial-size quantum state, in
such a way that the value of any one of those bits can later be proven with
the help of a polynomial-size quantum witness. \ We also show that any problem
in $\mathsf{QMA}$\ with polynomial-size quantum advice, is also in
$\mathsf{PSPACE}$\ with polynomial-size classical advice. \ This builds on our
earlier result that $\mathsf{BQP/qpoly}\subseteq\mathsf{PP/poly}$, and offers
an intriguing counterpoint to the recent discovery of Raz that
$\mathsf{QIP/qpoly}=\mathsf{ALL}$. \ Finally, we show that
$\mathsf{QCMA/qpoly}\subseteq\mathsf{PP/poly}$\ and that $\mathsf{QMA/rpoly}%
=\mathsf{QMA/poly}$.

\end{abstract}

\Section{Introduction\label{INTRO}}

Let Bob be a graduate student, and let $x$\ be an $n$-bit string representing
his thesis problem. \ Bob's goal is to learn $f\left(  x\right)  $, where
$f:\left\{  0,1\right\}  ^{n}\rightarrow\left\{  0,1\right\}  $ is a function
that maps every thesis problem to its binary answer (\textquotedblleft
yes\textquotedblright\ or \textquotedblleft no\textquotedblright). \ Bob knows
$x$ (his problem), but is completely ignorant of $f$ (how to solve the
problem). \ So to evaluate $f\left(  x\right)  $, he's going to need help from
his thesis advisor, Alice. \ Like most advisors, Alice is infinitely powerful,
wise, and benevolent. \ But also like most advisors, she's too busy to find
out what problems her students are working on. \ Instead, she just doles out
the same advice $s$ to all of them, which she hopes will let them evaluate
$f\left(  x\right)  $\ for any $x$ they might encounter. \ The question is,
how long does $s$\ have to be, for Bob to be able to evaluate $f\left(
x\right)  $\ for any $x$?

Clearly, the answer is that $s$ has to be $2^{n}$\ bits long---since otherwise
$s$ will underdetermine the truth table of $f$. \ Indeed, let $g\left(
x,s\right)  $ be Bob's best guess as to $f\left(  x\right)  $, given $x$ and
$s$. \ Then even if Alice can choose $s$\ probabilistically, and we only
require that $g\left(  x,s\right)  =f\left(  x\right)  $\ with probability at
least $2/3$\ for every $x$, still one can show that $s$ needs to be
$\Omega\left(  2^{n}\right)  $\ bits long.

But what if Alice is a quantum advisor, who can send Bob a quantum state
$\left\vert \psi_{f}\right\rangle $? \ Even in that case, Ambainis et
al.\ \cite{antv}\ showed that Alice has to send $\Omega\left(  2^{n}/n\right)
$\ qubits for Bob to succeed with probability at least $2/3$ on every $x$.
\ Subsequently Nayak \cite{nayak}\ improved this to $\Omega\left(
2^{n}\right)  $, meaning that there is no quantum improvement over the
classical bound. \ Since $2^{n}$\ qubits is too many for Alice to communicate
during her weekly meetings with Bob, it seems Bob is out of luck.

So in desperation, Bob turns for help to Merlin, the star student in his
department. \ Merlin knows $f$ as well as $x$,\ and can thus evaluate
$f\left(  x\right)  $. \ The trouble is that Merlin would prefer to take
credit for evaluating $f\left(  x\right)  $\ himself, so he might deliberately
mislead Bob. \ Furthermore, Merlin (whose brilliance is surpassed only by his
ego) insists that all communication with lesser students be one-way: Bob is to
listen in silence while Merlin lectures him.\ \ On the other hand, Merlin has
no time to give an exponentially long lecture, any more than Alice does.

With \textquotedblleft helpers\textquotedblright\ like these, Bob might ask,
who needs adversaries? \ And yet, is it possible that Bob could play Alice and
Merlin\ against each other---cross-checking Merlin's specific but unreliable
assertions against Alice's vague but reliable advice? \ In other words, does
there exist a randomized protocol satisfying the following properties?

\begin{enumerate}
\item[(i)] Alice and Merlin both send Bob $\operatorname*{poly}\left(
n\right)  $\ bits.

\item[(ii)] If Merlin tells Bob the truth about $f\left(  x\right)  $, then
there exists a message from Merlin that causes Bob to accept with probability
at least \thinspace$2/3$.

\item[(iii)] If Merlin lies about $f\left(  x\right)  $\ (i.e., claims that
$f\left(  x\right)  =1$\ when $f\left(  x\right)  =0$\ or vice versa), then no
message from Merlin causes Bob to accept with probability greater than $1/3$.
\end{enumerate}

It is relatively easy to show that the answer is no: if Alice sends $a$\ bits
to Bob and Merlin sends $w$ bits, then for Bob to succeed we must have
$a\left(  w+1\right)  =\Omega\left(  2^{n}\right)  $. \ Indeed, this is
basically tight: for all $w\geq1$, there exists a protocol in which Merlin
sends $w$\ bits and Alice sends $O\left(  \frac{2^{n}}{w}+n\right)  $\ bits.
\ Of course, even if Merlin didn't send anything, it would suffice for Alice
to send $2^{n}$\ bits. \ At the other extreme, if Merlin sends $2^{n}$\ bits,
then it suffices for Alice to send an $\Theta\left(  n\right)  $-bit
\textquotedblleft fingerprint\textquotedblright\ to authenticate Merlin's
message. \ But in any event, either Alice or Merlin will have to send an
exponentially-long message.

On the other hand, what if Alice and Merlin can both send \textit{quantum}
messages? \ Our main result will show that, even in this most general
scenario, \textit{Bob is still out of luck}. \ Indeed, if Alice sends $a$
qubits to Bob, and Merlin sends $w$ qubits, then Bob cannot succeed unless
$a\left(  w+1\right)  =\Omega\left(  2^{n}/n^{2}\right)  $. \ Apart from the
$n^{2}$\ factor\ (which we conjecture can be removed), this implies that no
quantum protocol is asymptotically better than the classical one. \ It
follows, then, that Bob ought to drop out of grad school and send his resume
to Google.

\SubSection{Banishing Merlin\label{DEMERLIN}}

But why should anyone care about this result, apart from Alice, Bob, Merlin,
and the Google recruiters? \ One reason is that the proof introduces a new
technique for removing existential quantifiers over quantum states, which
might be useful in other contexts. \ The basic idea is for Bob to loop over
all possible messages that Merlin could have sent, and accept if and only if
there exists a message that would cause him to accept. \ The problem is that
in the quantum case, the number of possible messages from Merlin is
doubly-exponential. \ So to loop over all of them, it seems we'd first need to
amplify Alice's message an exponential number of times. \ But surprisingly, we
show that this intuition is wrong:\ to account for any possible quantum
message from Merlin, it suffices to loop over all possible \textit{classical}
messages from Merlin! \ For, loosely speaking, any quantum state can
eventually be detected by the \textquotedblleft shadows\textquotedblright\ it
casts on computational basis states. \ However, turning this insight into a
\textquotedblleft de-Merlinization\textquotedblright\ procedure requires some
work: we need to amplify Alice's and Merlin's messages in a subtle way, and
then deal with the degradation of Alice's message that occurs regardless.

\SubSection{QMA With Quantum Advice\label{QMAINTRO}}

In any case, the main motivation for our result is that it implies a new
containment in quantum complexity theory: namely that%
\[
\mathsf{QMA/qpoly}\subseteq\mathsf{PSPACE/poly}.
\]
Here $\mathsf{QMA}$\ is the quantum version of $\mathsf{MA}$, and
$\mathsf{/qpoly}$\ means \textquotedblleft with polynomial-size quantum
advice.\textquotedblright\ \ Previously, it was not even known whether
$\mathsf{QMA/qpoly}=\mathsf{ALL}$, where $\mathsf{ALL}$\ is the class of all
languages! \ Nevertheless, some context might be helpful for understanding why
our new containment is of more than zoological interest.

Aaronson \cite{aar:adv}\ showed that $\mathsf{BQP/qpoly}\subseteq
\mathsf{PP/poly}$, where $\mathsf{BQP/qpoly}$\ is the class of problems
solvable in $\mathsf{BQP}$ with polynomial-size quantum advice. \ He also gave
an oracle relative to which $\mathsf{NP}\not \subset \mathsf{BQP/qpoly}$.
\ Together, these results seemed to place strong limits on the power of
quantum advice.

However, recently Raz \cite{raz:all} reopened the subject, by showing that in
some cases quantum advice can be extraordinarily powerful. \ In particular,
Raz showed that $\mathsf{QIP}\left(  2\right)  \mathsf{/qpoly}=\mathsf{ALL}$,
where $\mathsf{QIP}\left(  2\right)  $\ is the class of problems that admit
two-round quantum interactive proof systems. \ Raz's result was actually
foreshadowed by an observation in \cite{aar:adv}, that $\mathsf{PostBQP/qpoly}%
=\mathsf{ALL}$. \ Here $\mathsf{PostBQP}$\ is the class of problems solvable
in quantum polynomial time, if at any time we can measure the computer's state
and then \textquotedblleft postselect\textquotedblright\ on a particular
outcome occurring.\footnote{Here is the proof: given a Boolean function
$f:\left\{  0,1\right\}  ^{n}\rightarrow\left\{  0,1\right\}  $, take%
\[
\left\vert \psi_{n}\right\rangle =\frac{1}{2^{n/2}}\sum_{x\in\left\{
0,1\right\}  ^{n}}\left\vert x\right\rangle \left\vert f\left(  x\right)
\right\rangle
\]
as the advice. \ Then to evaluate $f\left(  x\right)  $\ on any $x$, simply
measure $\left\vert \psi_{n}\right\rangle $\ in the standard basis, and then
postselect on observing $\left\vert x\right\rangle $\ in the first register.}

These results should make any complexity theorist a little queasy, and not
only because jumping from $\mathsf{QIP}\left(  2\right)  $\ or
$\mathsf{PostBQP}$\ to $\mathsf{ALL}$\ is like jumping from a hilltop to the
edge of the universe. \ A more serious problem is that these results fail to
\textquotedblleft commute\textquotedblright\ with standard complexity
inclusions. \ For example, even though $\mathsf{PostBQP}$\ is strictly
contained in $\mathsf{BQEXPEXP}$, notice that $\mathsf{BQEXPEXP/qpoly}$\ is
(very) strictly contained in $\mathsf{PostBQP/qpoly}$!

\SubSection{The Quantum Advice Hypothesis\label{QAH}}

On the other hand, the same pathologies would occur with classical randomized
advice. \ For neither the result of Raz \cite{raz:all},\ nor that of Aaronson
\cite{aar:adv},\ makes any essential use of quantum mechanics. \ That is,
instead of saying that%
\[
\mathsf{QIP}\left(  2\right)  \mathsf{/qpoly}=\mathsf{PostBQP/qpoly}%
=\mathsf{ALL},
\]
we could equally well have said that%
\[
\mathsf{IP}\left(  2\right)  \mathsf{/rpoly}=\mathsf{PostBPP/rpoly}%
=\mathsf{ALL},
\]
where $\mathsf{IP}\left(  2\right)  $\ and $\mathsf{PostBPP}$\ are the
classical analogues of $\mathsf{QIP}\left(  2\right)  $\ and $\mathsf{PostBQP}%
$ respectively, and $\mathsf{/rpoly}$\ means \textquotedblleft with
polynomial-size randomized advice.\textquotedblright

Inspired by this observation, here we propose a general hypothesis: that
\textit{whenever quantum advice behaves like exponentially-long classical
advice, the reason has nothing to do with quantum mechanics}. \ More concretely:

\begin{itemize}
\item \textbf{The Quantum Advice Hypothesis:} For any \textquotedblleft
natural\textquotedblright\ complexity class $\mathcal{C}$, if $\mathcal{C}%
\mathsf{/qpoly}=\mathsf{ALL}$, then $\mathcal{C}\mathsf{/rpoly}=\mathsf{ALL}%
$\ as well.
\end{itemize}

The evidence for this hypothesis is simply that we have not been able to
refute it. \ In particular, in Appendix \ref{APPENDIX} we will show that
$\mathsf{QMA/rpoly}=\mathsf{QMA/poly}$. \ So if $\mathsf{QMA/qpoly}$ contained
all languages---which (at least to us)\ seemed entirely possible \textit{a
priori}---then we would have a clear counterexample to the hypothesis. \ In
our view, then, the significance of the $\mathsf{QMA/qpoly}\subseteq
\mathsf{PSPACE/poly}$\ result is that it confirms the quantum advice
hypothesis in the most nontrivial case considered so far.

To summarize, the quantum advice hypothesis has been confirmed for at least
four complexity classes: $\mathsf{BQP}$, $\mathsf{QMA}$, $\mathsf{PostBQP}$,
and $\mathsf{QIP}\left(  2\right)  $. \ It remains open for other classes,
such as $\mathsf{QMA}\left(  2\right)  $\ ($\mathsf{QMA}$ with two unentangled
yes-provers) and $\mathsf{QS}_{2}^{p}$\ ($\mathsf{QMA}$\ with competing
yes-prover and no-prover).

\SubSection{Outline of Paper\label{OUTLINE}}

\begin{itemize}
\item Section \ref{PRELIM} surveys the complexity classes, communication
complexity measures, and quantum information notions used in this paper.

\item Section \ref{APP} states our \textquotedblleft De-Merlinization
Theorem,\textquotedblright\ and then proves three of its implications: (i) a
lower bound on the QMA communication complexity\ of random access coding, (ii)
a general lower bound on QMA\ communication complexity, and (iii) the
inclusion $\mathsf{QMA/qpoly}\subseteq\mathsf{PSPACE/poly}$.

\item Section \ref{PROOF} proves the De-Merlinization Theorem itself.

\item Section \ref{OPEN} concludes with some open problems.

\item Appendix \ref{APPENDIX} proves a few other complexity results, including
$\mathsf{QCMA/qpoly}\subseteq\mathsf{PP/poly}$\ and $\mathsf{QMA/rpoly}%
=\mathsf{QMA/poly}$.
\end{itemize}

\Section{Preliminaries\label{PRELIM}}

\SubSection{Complexity Classes\label{ZOOPRELIM}}

We assume familiarity with standard complexity classes like $\mathsf{BQP}$,
$\mathsf{P/poly}$, and $\mathsf{MA}$. \ The class $\mathsf{QMA}$ (Quantum
Merlin-Arthur) consists of all languages for which a `yes' answer can be
verified in quantum polynomial time, given a polynomial-size quantum witness
state $\left\vert \varphi\right\rangle $. \ The completeness and soundness
errors are $1/3$. \ The class $\mathsf{QCMA}$\ (Quantum Classical
Merlin-Arthur) is the same as $\mathsf{QMA}$, except that now the witness must
be classical. \ It is not known whether $\mathsf{QMA}=\mathsf{QCMA}$. \ See
the Complexity Zoo\footnote{http://qwiki.caltech.edu/wiki/Complexity\_Zoo} for
more information about these and other classes.

Given a complexity class $\mathcal{C}$, we write $\mathcal{C}\mathsf{/poly}$,
$\mathcal{C}\mathsf{/rpoly}$, and $\mathcal{C}\mathsf{/qpoly}$\ to denote
$\mathcal{C}$\ with polynomial-size deterministic, randomized, and quantum
advice respectively.\footnote{We can also write $\mathcal{C}\mathsf{/rl{}og}%
$\ (for $\mathcal{C}$\ with logarithmic-size randomized advice),
$\mathcal{C}\mathsf{/ql{}og}$, and so on.} \ So for example,
$\mathsf{BPP/rpoly}$\ is the class of languages decidable by a $\mathsf{BPP}$
machine, given a sample from a distribution\ $\mathcal{D}_{n}$\ over
polynomial-size advice strings which depends only on the input length $n$.
\ It is clear that $\mathsf{BPP/rpoly}=\mathsf{BPP/poly}=\mathsf{P/poly}$.
\ However, in other cases the statement $\mathcal{C}\mathsf{/rpoly}%
=\mathcal{C}\mathsf{/poly}$\ is harder to prove or is even false.

Admittedly, the $\mathsf{/rpoly}$\ and $\mathsf{/qpoly}$\ operators are not
always well-defined: for example, $\mathsf{P/qpoly}$\ is just silly, and
$\mathsf{AM/rpoly}$\ seems ambiguous (since who gets to sample from the advice
distribution?). \ For interactive proof classes, the general rule we adopt is
that \textit{only the verifier gets to \textquotedblleft
measure\textquotedblright\ the advice}. \ In other words, the prover (or
provers) knows the advice distribution $\mathcal{D}_{n}$\ or advice state
$\left\vert \psi_{n}\right\rangle $, but not the actual results of sampling
from $\mathcal{D}_{n}$\ or measuring $\left\vert \psi_{n}\right\rangle $. \ In
the case of $\mathsf{/rpoly}$, the justification for this rule is that, if the
prover knew the sample from $\mathcal{D}_{n}$, then we would immediately get
$\mathcal{C}\mathsf{/rpoly}=\mathcal{C}\mathsf{/poly}$\ for all interactive
proof classes $\mathcal{C}$, which is too boring. \ In the case of
$\mathsf{/qpoly}$, the justification is that the verifier should be allowed to
measure $\left\vert \psi_{n}\right\rangle $\ at any time and in any basis it
likes, and it seems perverse to require the results of such measurements to be
relayed instantly to the prover.

In a private-coin protocol, the verifier might choose to reveal some or all of
the measurement results to the prover, but in a public-coin protocol, the
verifier must send a uniform random message that is uncorrelated with the
advice. \ Indeed, this explains how it can be true that $\mathsf{IP}\left(
2\right)  \mathsf{/rpoly}\neq\mathsf{AM/rpoly}$ (the former equals
$\mathsf{ALL}$, while the latter equals\ $\mathsf{NP/poly}$), even though
Goldwasser and Sipser \cite{gs}\ famously showed that $\mathsf{IP}\left(
2\right)  =\mathsf{AM}$\ in the uniform setting.

For the complexity classes $\mathcal{C}$\ that appear in this paper, it should
generally be obvious what we mean by $\mathcal{C}\mathsf{/rpoly}$\ or
$\mathcal{C}\mathsf{/qpoly}$. \ But to fix ideas, let us now formally define
$\mathsf{QMA/qpoly}$.

\begin{definition}
$\mathsf{QMA/qpoly}$\ is the class of languages $L\subseteq\left\{
0,1\right\}  ^{\ast}$\ for which there exists a polynomial-time quantum
verifier $\mathcal{Q}$, together with quantum advice states $\left\{
\left\vert \psi_{n}\right\rangle \right\}  _{n\geq1}$, such that for all
$x\in\left\{  0,1\right\}  ^{n}$:

\begin{enumerate}
\item[(i)] If $x\in L$, then there exists a quantum witness $\left\vert
\varphi\right\rangle $\ such that $\mathcal{Q}$\ accepts with probability at
least $2/3$\ given $\left\vert x\right\rangle \left\vert \psi_{n}\right\rangle
\left\vert \varphi\right\rangle $\ as input.

\item[(ii)] If $x\notin L$, then for all pure states\footnote{By linearity,
this is equivalent to quantifying over all mixed states of the witness
register.} $\left\vert \varphi\right\rangle $ of the witness register,
$\mathcal{Q}$\ accepts with probability at most $1/3$\ given $\left\vert
x\right\rangle \left\vert \psi_{n}\right\rangle \left\vert \varphi
\right\rangle $\ as input.
\end{enumerate}

Here $\left\vert \psi_{n}\right\rangle $\ and $\left\vert \varphi\right\rangle
$\ both consist of $p\left(  n\right)  $\ qubits for some fixed polynomial
$p$. \ Also, $\mathcal{Q}$\ can accept with arbitrary probability\ if given a
state other than $\left\vert \psi_{n}\right\rangle $\ in the advice register.
\end{definition}

One other complexity class we will need is $\mathsf{PostBQP}$, or
$\mathsf{BQP}$ with postselection.

\begin{definition}
$\mathsf{PostBQP}$ is the class of languages $L\subseteq\left\{  0,1\right\}
^{\ast}$\ for which there exists a polynomial-time quantum algorithm such that
for all $x\in\left\{  0,1\right\}  ^{n}$, when the algorithm terminates:

\begin{enumerate}
\item[(i)] The first qubit is $\left\vert 1\right\rangle $\ with nonzero probability.

\item[(ii)] If $x\in L$, then conditioned on the first qubit being $\left\vert
1\right\rangle $, the second qubit is $\left\vert 1\right\rangle $\ with
probability at least $2/3$.

\item[(iii)] If $x\notin L$, then conditioned on the first qubit being
$\left\vert 1\right\rangle $, the second qubit is $\left\vert 1\right\rangle
$\ with probability at most $1/3$.
\end{enumerate}
\end{definition}

One can similarly define $\mathsf{PostBQPSPACE}$, $\mathsf{PostBQEXP}$, and so
on. \ We will use a result of Aaronson \cite{aar:pp}, which characterizes
$\mathsf{PostBQP}$\ as simply the classical complexity class $\mathsf{PP}$.

\SubSection{Communication Complexity\label{CCPRELIM}}

Let $f:\left\{  0,1\right\}  ^{N}\times\left\{  0,1\right\}  ^{M}%
\rightarrow\left\{  0,1\right\}  $\ be a Boolean function.\ \ Suppose Alice
has an $N$-bit string $X$ and Bob has an $M$-bit string $Y$. \ Then
$\operatorname*{D}^{1}\left(  f\right)  $ is the deterministic one-way
communication complexity of $f$: that is, the minimum number of bits that
Alice must send to Bob, for Bob to be able to output $f\left(  X,Y\right)
$\ with certainty for any $\left(  X,Y\right)  $\ pair. \ If we let Alice's
messages be randomized, and only require Bob to be correct with probability
$2/3$, then we obtain $\operatorname*{R}^{1}\left(  f\right)  $, the
bounded-error randomized one-way communication complexity of $f$. \ Finally,
if we let Alice's messages be quantum, then we obtain $\operatorname*{Q}%
\nolimits^{1}\left(  f\right)  $, the bounded-error quantum one-way
communication complexity of $f$.\footnote{We assume no shared randomness or
entanglement. \ Also, we assume for simplicity that Alice can only send pure
states; note that this increases the message length by at most a
multiplicative factor of $2$ (or an additive factor of $\log N$,\ if we use
Newman's Theorem \cite{newman}).} \ Clearly $\operatorname*{Q}\nolimits^{1}%
\left(  f\right)  \leq\operatorname*{R}^{1}\left(  f\right)  \leq
\operatorname*{D}^{1}\left(  f\right)  $\ for all $f$. \ See Klauck
\cite{klauck:cc}\ for more detailed definitions of these measures.

Now suppose that, in addition to a quantum message $\left\vert \psi
_{X}\right\rangle $\ from Alice, Bob also receives a quantum witness
$\left\vert \varphi\right\rangle $\ from Merlin, whose goal is to convince Bob
that $f\left(  X,Y\right)  =1$.\footnote{For convenience, from now on we
assume that Merlin only needs to prove statements of the form $f\left(
X,Y\right)  =1$, not $f\left(  X,Y\right)  =0$. \ For our actual results, it
will make no difference whether we adopt this assumption (corresponding to the
class $\mathsf{QMA}$), or the assumption in Section \ref{INTRO}%
\ (corresponding to $\mathsf{QMA\cap coQMA}$).} \ We say Alice and Bob
\textit{succeed} if for all $X,Y$,

\begin{enumerate}
\item[(i)] If $f\left(  X,Y\right)  =1$, then there exists a $\left\vert
\varphi\right\rangle $ such that Bob accepts $\left\vert Y\right\rangle
\left\vert \psi_{X}\right\rangle \left\vert \varphi\right\rangle $\ with
probability at least $2/3$.

\item[(ii)] If $f\left(  X,Y\right)  =0$, then for all $\left\vert
\varphi\right\rangle $, Bob accepts $\left\vert Y\right\rangle \left\vert
\psi_{X}\right\rangle \left\vert \varphi\right\rangle $\ with probability at
most $1/3$.
\end{enumerate}

Call a protocol \textquotedblleft$\left(  a,w\right)  $\textquotedblright\ if
Alice's message consists of $a$ qubits and Merlin's consists of $w$ qubits.
\ Then for all integers $w\geq0$, we let $\operatorname*{QMA}\nolimits_{w}%
^{1}\left(  f\right)  $\ denote the \textquotedblleft$\operatorname*{QMA}%
\nolimits_{w}$ one-way\ communication complexity\textquotedblright\ of $f$:
that is, the minimum $a$ for which there exists an $\left(  a,w\right)
$\ protocol\ such that Alice and Bob succeed. \ Clearly $\operatorname*{QMA}%
\nolimits_{w}^{1}\left(  f\right)  \leq\operatorname*{Q}\nolimits^{1}\left(
f\right)  $, with equality when $w=0$.

\SubSection{Quantum Information\label{QIPRELIM}}

Here we review some basic facts about mixed states. \ Further details can be
found in Nielsen and Chuang \cite{nc} for example.

Given two mixed states $\rho$\ and $\sigma$, the \textit{fidelity} $F\left(
\rho,\sigma\right)  $ is the maximum possible value of $\left\langle
\psi|\varphi\right\rangle $, where $\left\vert \psi\right\rangle $\ and
$\left\vert \varphi\right\rangle $\ are purifications of $\rho$\ and $\sigma$
respectively. \ Also, given a measurement $M$, let $\mathcal{D}_{M}\left(
\rho\right)  $\ be the probability distribution over measurement outcomes if
$M$ is applied to $\rho$. \ Then the \textit{trace distance} $\left\Vert
\rho-\sigma\right\Vert _{\operatorname*{tr}}$\ equals the maximum, over all
possible measurements $M$, of $\left\Vert \mathcal{D}_{M}\left(  \rho\right)
-\mathcal{D}_{M}\left(  \sigma\right)  \right\Vert $,\ where%
\[
\left\Vert \mathcal{D}-\mathcal{D}^{\prime}\right\Vert =\frac{1}{2}\sum
_{i=1}^{N}\left\vert p_{i}-p_{i}^{\prime}\right\vert
\]
is the variation distance between $\mathcal{D}=\left(  p_{1},\ldots
,p_{N}\right)  $\ and $\mathcal{D}^{\prime}=\left(  p_{1}^{\prime}%
,\ldots,p_{N}^{\prime}\right)  $. \ For all $\rho$\ and $\sigma$, we have the
following relation between fidelity and trace distance:%
\[
\left\Vert \rho-\sigma\right\Vert _{\operatorname*{tr}}\leq\sqrt{1-F\left(
\rho,\sigma\right)  ^{2}}.
\]

Throughout this paper, we use $\mathcal{H}_{N}$\ to denote $N$-dimensional
Hilbert space. \ One fact we will invoke repeatedly is that, if $I$\ is the
maximally mixed state in $\mathcal{H}_{N}$, then%
\[
I=\frac{1}{N}\sum_{j=1}^{N}\left\vert \psi_{j}\right\rangle \left\langle
\psi_{j}\right\vert
\]
where $\left\{  \left\vert \psi_{1}\right\rangle ,\ldots,\left\vert \psi
_{N}\right\rangle \right\}  $\ is \textit{any} orthonormal basis\ for
$\mathcal{H}_{N}$.

\Section{De-Merlinization and Its Applications\label{APP}}

Our main result, the \textquotedblleft De-Merlinization
Theorem,\textquotedblright\ allows us to lower-bound $\operatorname*{QMA}%
\nolimits_{w}^{1}\left(  f\right)  $\ in terms of the ordinary quantum
communication complexity\ $\operatorname*{Q}\nolimits^{1}\left(  f\right)  $.
\ In this section we state the theorem and derive its implications for random
access coding (in Section \ref{RAC}), one-way communication complexity (in
Section \ref{1WAY}), and complexity theory (in Section \ref{QMAPP}). \ The
theorem itself will be proved in Section \ref{PROOF}.

\begin{theorem}
[De-Merlinization Theorem]\label{demerlin}For all Boolean functions
$f$\ (partial or total) and all $w\geq2$,%
\[
\operatorname*{Q}\nolimits^{1}\left(  f\right)  =O\left(  \operatorname*{QMA}%
\nolimits_{w}^{1}\left(  f\right)  \cdot w\log^{2}w\right)  .
\]
Furthermore, given an algorithm for the $\operatorname*{QMA}\nolimits_{w}^{1}%
$\ protocol, Bob can efficiently generate an algorithm for the
$\operatorname*{Q}\nolimits^{1}$\ protocol. \ If the former uses $C$ gates and
$S$\ qubits of memory, then the latter uses $C\cdot S^{O\left(  S\right)  }%
$\ gates and $O\left(  S^{2}\log^{2}S\right)  $\ qubits of memory.
\end{theorem}

\SubSection{Application I: Random Access Coding\label{RAC}}

Following Ambainis et al.\ \cite{antv}, let us define the \textit{random
access coding} (or $\operatorname*{RAC}$) problem as follows. \ Alice has an
$N$-bit string $X=x_{1}\ldots x_{N}$ and Bob has an index $i\in\left\{
1,\ldots,N\right\}  $. \ The players' goal is for Bob to learn $x_{i}$.

In our setting, Bob receives not only an $a$-bit message from Alice, but also
a $w$-bit message from Merlin. \ If $x_{i}=1$, then there should exist a
message from Merlin that causes Bob to accept with probability at least
$2/3$;\ while if $x_{i}=0$,\ then no message from Merlin should cause Bob to
accept with probability greater than $1/3$. \ We are interested in the minimum
$a,w$\ for which Alice and Bob can succeed.

For completeness, before stating our results for the quantum case, let us
first pin down the classical case---that is, the case in which Alice and
Merlin both send classical messages, and Alice's message can be randomized.
\ Obviously, if Merlin sends $0$\ bits, then\ Alice needs to send
$\Theta\left(  N\right)  $\ bits; this is just the ordinary
$\operatorname*{RAC}$ problem studied by Ambainis et al.\ \cite{antv}. \ At
the other extreme, if Merlin sends the $N$-bit message $X$, then it suffices
for Alice to send an $O\left(  \log N\right)  $-bit fingerprint of $X$. \ For
intermediate message lengths, we can interpolate between these two extremes.

\begin{theorem}
\label{protocol}For all $a,w$ such that $aw\geq N$, there exists a randomized
$\left(  a+O\left(  \log N\right)  ,w\right)  $\ protocol for RAC---that is, a
protocol in which Alice sends $a+O\left(  \log N\right)  $\ bits and Merlin
sends $w$\ bits.
\end{theorem}

\begin{proof}
The protocol is as follows: first Alice divides her string $X=x_{1}\ldots
x_{N}$\ into $a$\ substrings $Y_{1},\ldots,Y_{a}$, each at most $w$ bits long.
\ She then maps each $Y_{j}$\ to an encoded substring $Y_{j}^{\prime}=g\left(
Y_{j}\right)  $, where $g:\left\{  0,1\right\}  ^{w}\rightarrow\left\{
0,1\right\}  ^{W}$\ is a constant-rate error-correcting code satisfying
$W=O\left(  w\right)  $. \ Next she chooses $k\in\left\{  1,\ldots,W\right\}
$\ uniformly at random. \ Finally, she sends Bob $k$ (which requires $O\left(
\log N\right)  $\ bits of communication), together with the $k^{th}$\ bit of
$Y_{j}^{\prime}$ for every $j\in\left\{  1,\ldots,a\right\}  $.

Now if Merlin is honest, then he sends Bob the substring $Y_{j}\in\left\{
0,1\right\}  ^{w}$\ of $X$ containing the $x_{i}$\ that Bob is interested in.
\ This allows Bob to learn $x_{i}$. \ Furthermore, if Merlin cheats by sending
some $Y\neq Y_{j}$, then Bob can detect this with constant probability, by
cross-checking the $k^{th}$\ bit of\ $g\left(  Y\right)  $ against the
$k^{th}$\ bit of $Y_{j}^{\prime}$\ as sent by Alice.
\end{proof}

Using a straightforward amplification trick, we can show that the protocol of
Theorem \ref{protocol}\ is essentially optimal.

\begin{theorem}
\label{tight}If there exists a randomized $\left(  a,w\right)  $\ protocol for
RAC, then $a\left(  w+1\right)  =\Omega\left(  N\right)  $ and $a=\Omega
\left(  \log N\right)  $.
\end{theorem}

\begin{proof}
We first show that $a\left(  w+1\right)  =\Omega\left(  N\right)  $. \ First
Alice amplifies her message to Bob by sending $W=O\left(  w+1\right)  $
independent copies of it. \ For any fixed message of Merlin, this reduces
Bob's error probability to at most (say) $2^{-2\left(  w+1\right)  }$. \ So
now Bob can ignore Merlin, and loop over all $2^{w}$\ messages $z\in\left\{
0,1\right\}  ^{w}$\ that Merlin \textit{could} have sent, accepting if and
only if there exists a $z$ that would cause him to accept. \ This yields an
ordinary protocol for the RAC problem in which Alice sends $aW$\ bits to Bob.
\ But Ambainis et al.\ \cite{antv} showed that any such protocol requires
$\Omega\left(  N\right)  $\ bits; hence $a\left(  w+1\right)  =\Omega\left(
N\right)  $.

That Alice needs to send $\Omega\left(  \log N\right)  $\ bits follows by a
simple counting argument: let $\mathcal{D}_{X}$\ be Alice's message
distribution given an input $X$. \ Then $\mathcal{D}_{X}$\ and $\mathcal{D}%
_{Y}$\ must have constant\ variation distance for all $X\neq Y$, if Bob is to
distinguish $X$ from $Y$ with constant bias.
\end{proof}

Together, Theorems \ref{protocol} and \ref{tight}\ provide the complete story
for the classical case, up to a constant factor. \ In the quantum case, the
situation is no longer so simple, but we can give a bound that is tight up to
a polylog factor.

\begin{theorem}
\label{qraclb}If there exists a quantum $\left(  a,w\right)  $ protocol for
RAC, then%
\[
a\left(  w+1\right)  =\Omega\left(  \frac{N}{\log^{2}N}\right)  .
\]

\end{theorem}

\begin{proof}
If $w=0$\ or $w=1$\ then clearly $a=\Omega\left(  N\right)  $,\ so assume
$w\geq2$. By Theorem \ref{demerlin},%
\begin{align*}
\operatorname*{Q}\nolimits^{1}\left(  \operatorname*{RAC}\right)   &
=O\left(  \operatorname*{QMA}\nolimits_{w}^{1}\left(  \operatorname*{RAC}%
\right)  \cdot w\log^{2}w\right) \\
&  =O\left(  aw\cdot\log^{2}N\right)  .
\end{align*}
But Nayak \cite{nayak}\ showed that $\operatorname*{Q}\nolimits^{1}\left(
\operatorname*{RAC}\right)  =\Omega\left(  N\right)  $, and hence
$aw=\Omega\left(  N/\log^{2}N\right)  $.
\end{proof}

Clearly Theorem \ref{qraclb}\ can be improved when $w$ is very small or very
large. \ For when $w=0$, we have $a=\Omega\left(  N\right)  $; while for any
$w$, a simple counting argument (as in the classical case) yields
$a=\Omega\left(  \log N\right)  $. \ We believe that Theorem \ref{qraclb}\ can
be improved for intermediate $w$ as well, since we do not know of any quantum
protocol that beats the classical protocol of Theorem \ref{protocol}.

\SubSection{Application II: One-Way Communication\label{1WAY}}

Theorem \ref{demerlin}\ yields lower bounds on QMA communication complexity,
not only for the random access coding problem, but for other problems as well.
\ For Aaronson \cite{aar:adv}\ showed the following general relationship
between $\operatorname*{D}\nolimits^{1}\left(  f\right)  $\ and
$\operatorname*{Q}\nolimits_{2}^{1}\left(  f\right)  $:

\begin{theorem}
[\cite{aar:adv}]\label{dqthm}For all Boolean functions $f:\left\{
0,1\right\}  ^{N}\times\left\{  0,1\right\}  ^{M}\rightarrow\left\{
0,1\right\}  $\ (partial or total),%
\[
\operatorname*{D}\nolimits^{1}\left(  f\right)  =O\left(  M\operatorname*{Q}%
\nolimits_{2}^{1}\left(  f\right)  \log\operatorname*{Q}\nolimits_{2}%
^{1}\left(  f\right)  \right)  .
\]

\end{theorem}

Combining Theorem \ref{dqthm} with Theorem \ref{demerlin}, we obtain the
following relationship between $\operatorname*{D}\nolimits^{1}\left(
f\right)  $\ and $\operatorname*{QMA}\nolimits_{w}^{1}\left(  f\right)  $.
\ For all $f:\left\{  0,1\right\}  ^{N}\times\left\{  0,1\right\}
^{M}\rightarrow\left\{  0,1\right\}  $\ (partial or total) and all $w\geq2$,%
\[
\operatorname*{D}\nolimits^{1}\left(  f\right)  =O\left(  M\cdot w\log
^{3}w\cdot\operatorname*{QMA}\nolimits_{w}^{1}\left(  f\right)  \log
\operatorname*{QMA}\nolimits_{w}^{1}\left(  f\right)  \right)  .
\]

\SubSection{\label{QMAPP}Application III: Upper-Bounding QMA/qpoly}

We now explain why the containment\ $\mathsf{QMA/qpoly}\subseteq
\mathsf{PSPACE/poly}$\ follows from the De-Merlinization Theorem. \ The first
step is to observe a weaker result that follows from that theorem:

\begin{lemma}
\label{bqpspacethm}$\mathsf{QMA/qpoly}\subseteq\mathsf{BQPSPACE/qpoly}$.
\end{lemma}

\begin{proof}
Given a language $L\in\mathsf{QMA/qpoly}$, let $L_{n}:\left\{  0,1\right\}
^{n}\rightarrow\left\{  0,1\right\}  $\ be the Boolean function defined by
$L_{n}\left(  x\right)  =1$\ if $x\in L$\ and $L_{n}\left(  x\right)
=0$\ otherwise. \ Then if we interpret Alice's input as the truth table of
$L_{n}$, Bob's input as $x$, and $S$ as the number of qubits used by the
$\mathsf{QMA/qpoly}$ machine, the lemma follows immediately from Theorem
\ref{demerlin}.
\end{proof}

Na\"{\i}vely, Lemma \ref{bqpspacethm} might seem obvious, since it is
well-known that $\mathsf{QMA}\subseteq\mathsf{PSPACE}$. \ But remember that
even if $\mathcal{C}\subseteq\mathcal{D}$, it need not follow that
$\mathcal{C}\mathsf{/qpoly}\subseteq\mathcal{D}\mathsf{/qpoly}$.

The next step is to replace the quantum advice by classical advice.

\begin{lemma}
$\mathsf{BQPSPACE/qpoly}\subseteq\mathsf{PostBQPSPACE/poly}$.
\end{lemma}

\begin{proof}
Follows from the same argument used by Aaronson \cite{aar:adv}\ to show that
$\mathsf{BQP/qpoly}\subseteq\mathsf{PostBQP/poly}$. \ All we need to do is
replace polynomial time by polynomial space.
\end{proof}

Finally, we observe a simple generalization of Watrous's theorem
\cite{watrous:space} that $\mathsf{BQPSPACE}=\mathsf{PSPACE}$.

\begin{lemma}
\label{postbqpspace}$\mathsf{PostBQPSPACE}=\mathsf{PSPACE}$.
\end{lemma}

\begin{proof}
[Proof Sketch]Ladner \cite{ladner:ppspace}\ showed that $\mathsf{PPSPACE}%
=\mathsf{PSPACE}$. \ Intuitively, given the computation graph of a
$\mathsf{PPSPACE}$\ machine, we want to decide in $\mathsf{PSPACE}$\ whether
the number of accepting paths exceeds the number of rejecting paths. \ To do
so we use divide-and-conquer, as in the proof of Savitch's theorem\ that
$\mathsf{NPSPACE}=\mathsf{PSPACE}$. \ An obvious difficulty is that the
numbers of paths could be \textit{doubly} exponential, and therefore take
exponentially many bits to store. \ But we can deal with that by computing
each bit of the numbers separately. \ Here we use the fact that there exist
$\mathsf{NC}^{1}$\ circuits for addition, and hence\ addition of $2^{n}%
$-bit\ integers is \textquotedblleft locally\textquotedblright\ in
$\mathsf{PSPACE}$.

If each path is weighted by a complex amplitude, then it is easy to see that
the same idea lets us sum the amplitudes over all paths. \ We can thereby
simulate $\mathsf{BQPSPACE}$\ and $\mathsf{PostBQPSPACE}$\ in $\mathsf{PSPACE}%
$\ as well.
\end{proof}

In particular, Lemma \ref{postbqpspace}\ implies that
$\mathsf{PostBQPSPACE/poly}=\mathsf{PSPACE/poly}$. \ (For note that unlike
randomized and quantum advice, deterministic advice commutes with standard
complexity class inclusions.)

Putting it all together, we obtain:

\begin{theorem}
\label{pspacethm}$\mathsf{QMA/qpoly}\subseteq\mathsf{PSPACE/poly}$.
\end{theorem}

As a final remark, let $\mathsf{QAM}$\ be the quantum analogue of
$\mathsf{AM}$, in which Arthur sends a public random string to Merlin, and
then Merlin responds with a quantum state. Marriott and Watrous \cite{mw}
observed that$\ \mathsf{QAM}=\mathsf{BP}\mathsf{\cdot QMA}$. \ So%
\[
\mathsf{QAM/qpoly}=\mathsf{BP}\mathsf{\cdot QMA/qpoly}=\mathsf{QMA/qpoly}%
\text{,}%
\]
since we can hardwire the random string into the quantum advice. \ Hence
$\mathsf{QAM/qpoly}\subseteq\mathsf{PSPACE/poly}$\ as well. \ This offers an
interesting contrast with the result of Raz \cite{raz:all}\ that
$\mathsf{QIP}\left(  2\right)  \mathsf{/qpoly}=\mathsf{ALL}$.

\Section{Proof of The De-Merlinization Theorem\label{PROOF}}

We now proceed to the proof of Theorem \ref{demerlin}. \ In Section
\ref{QIL}\ we prove several lemmas about damage to quantum states, and in
particular, the effect of the damage caused by earlier measurements of a state
on the outcomes of later measurements. \ Section \ref{AMP}\ then gives our
procedure for amplifying Bob's error probability, after explaining why the
more obvious procedures fail. \ Finally, Section \ref{MAIN}\ puts together the pieces.

\SubSection{Quantum Information Lemmas\label{QIL}}

In this section we prove several lemmas that will be needed for the main
result. \ The first lemma is a simple variant of Lemma 2.2\ from
\cite{aar:adv}; we include a proof for completeness.

\begin{lemma}
[Almost As Good As New Lemma]\label{goodasnew}Suppose a $2$-outcome POVM
measurement of a mixed state $\rho$\ yields outcome $1$ with probability
$\varepsilon$. \ Then after the measurement, and assuming outcome $0$ is
observed, we obtain a new state $\rho_{0}$ such that $\left\Vert \rho-\rho
_{0}\right\Vert _{\operatorname*{tr}}\leq\sqrt{\varepsilon}$.
\end{lemma}

\begin{proof}
Let $\left\vert \psi\right\rangle $\ be a purification of $\rho$. Then we can
write\ $\left\vert \psi\right\rangle $ as $\sqrt{1-\varepsilon}\left\vert
\psi_{0}\right\rangle +\sqrt{\varepsilon}\left\vert \psi_{1}\right\rangle $,
where $\left\vert \psi_{0}\right\rangle $ is a purification of $\rho_{0}$ and
$\left\langle \psi_{0}|\psi_{1}\right\rangle =0$. \ So the fidelity
between\ $\rho$\ and $\rho_{0}$\ is%
\[
F\left(  \rho,\rho_{0}\right)  \geq\left\langle \psi|\psi_{0}\right\rangle
=\sqrt{1-\varepsilon}.
\]
Therefore%
\[
\left\Vert \rho-\rho_{0}\right\Vert _{\operatorname*{tr}}\leq\sqrt{1-F\left(
\rho,\rho_{0}\right)  ^{2}}\leq\sqrt{\varepsilon}.
\]

\end{proof}

The next lemma, which we call the \textquotedblleft quantum union
bound,\textquotedblright\ abstracts one of the main ideas from \cite{antv}.

\begin{lemma}
[Quantum Union Bound]\label{unionbound}Let $\rho$\ be a mixed state, and let
$\left\{  \Lambda_{1},\ldots,\Lambda_{T}\right\}  $\ be a set of $2$-outcome
POVM measurements. \ Suppose each $\Lambda_{t}$\ yields outcome $1$ with
probability at most $\varepsilon$\ when applied to $\rho$. \ Then if we apply
$\Lambda_{1},\ldots,\Lambda_{T}$\ in sequence to $\rho$, the probability that
at least one of these measurements yields outcome $1$ is at most
$T\sqrt{\varepsilon}$.
\end{lemma}

\begin{proof}
Follows from a hybrid argument, almost identical to Claim 4.1 of Ambainis et
al.\ \cite{antv}. \ More explicitly, by the principle of deferred measurement,
we can replace each measurement $\Lambda_{t}$\ by a unitary $U_{t}$\ that
CNOT's the measurement outcome into an ancilla qubit. \ Let $\rho_{0}%
=\rho\otimes\left\vert 0\cdots0\right\rangle \left\langle 0\cdots0\right\vert
$ be the initial state of the system plus $T$ ancilla qubits. \ Then by the
same idea as in Lemma \ref{goodasnew}, for all $t$ we have%
\[
\left\Vert U_{t}\rho_{0}U_{t}^{-1}-\rho_{0}\right\Vert _{\operatorname*{tr}%
}\leq\sqrt{\varepsilon}.
\]
So letting%
\[
\rho_{t}:=U_{T}\cdots U_{T-t+1}\rho_{0}U_{T-t+1}^{-1}\cdots U_{T}^{-1},
\]
by unitarity we also have%
\begin{align*}
\left\Vert \rho_{t+1}-\rho_{t}\right\Vert _{\operatorname*{tr}}  &
=\left\Vert
\begin{array}
[c]{l}%
U_{T}\cdots U_{T-t}\rho_{0}U_{T-t}^{-1}\cdots U_{T}^{-1}\\
-U_{T}\cdots U_{T-t+1}\rho_{0}U_{T-t+1}^{-1}\cdots U_{T}^{-1}%
\end{array}
\right\Vert _{\operatorname*{tr}}\\
&  =\left\Vert U_{T-t}\rho_{0}U_{T-t}^{-1}-\rho_{0}\right\Vert
_{\operatorname*{tr}}\\
&  \leq\sqrt{\varepsilon},
\end{align*}
and hence $\left\Vert \rho_{T}-\rho_{0}\right\Vert _{\operatorname*{tr}}\leq
T\sqrt{\varepsilon}$\ by the triangle inequality.

Now let $M$ be a measurement that returns the logical OR of the $T$ ancilla
qubits, and let $\mathcal{D}\left(  \rho\right)  $\ be the distribution over
the outcomes ($0$ and $1$) when $M$ is applied to $\rho$. \ Suppose
$M$\ yields outcome $1$ with probability $p$ when applied to $\rho_{T}$.
\ Then since $M$ yields outcome $1$ with probability $0$ when applied to
$\rho_{0}$, the variation distance $\left\Vert \mathcal{D}\left(  \rho
_{T}\right)  -\mathcal{D}\left(  \rho_{0}\right)  \right\Vert $\ is equal to
$p$. \ So by the definition of trace distance,%
\[
p\leq\left\Vert \rho_{T}-\rho_{0}\right\Vert _{\operatorname*{tr}}\leq
T\sqrt{\varepsilon}.
\]

\end{proof}

Finally, we give a lemma that is key to our result. \ This lemma, which we
call the \textquotedblleft quantum OR bound,\textquotedblright\ is a sort of
converse to the quantum union bound. \ It says that, for all quantum circuits
$\Lambda$ and advice states $\left\vert \psi\right\rangle $, if there exists a
witness state $\left\vert \varphi\right\rangle $\ such that $\Lambda$\ accepts
$\left\vert \psi\right\rangle \left\vert \varphi\right\rangle $\ with high
probability, then we can also cause $\Lambda$\ to accept with high probability
by repeatedly running $\Lambda$ on $\left\vert \psi\right\rangle \left\vert
j\right\rangle $,\ where $\left\vert j\right\rangle $\ is a random basis state
of the witness register, and then taking the logical OR of the outcomes. \ One
might worry that, as we run $\Lambda$\ with various $\left\vert j\right\rangle
$'s, the state of the advice register might become corrupted to something far
from $\left\vert \psi\right\rangle $. \ However, we show that if this happens,
then it can only be because one of the measurements has already accepted with
high probability.

\begin{lemma}
[Quantum OR Bound]\label{orbound}Let $\Lambda$\ be a $2$-outcome POVM
measurement on a bipartite Hilbert space $\mathcal{H}_{A}\otimes
\mathcal{H}_{B}$. \ Also, let $\left\{  \left\vert 1\right\rangle
,\ldots,\left\vert N\right\rangle \right\}  $\ be any orthonormal basis for
$\mathcal{H}_{B}$, and for all $j\in\left\{  1,\ldots,N\right\}  $, let
$\Lambda_{j}$\ be the POVM on $\mathcal{H}_{A}$\ induced by applying $\Lambda
$\ to $\mathcal{H}_{A}\otimes\left\vert j\right\rangle $. \ Suppose there
exists a product state $\rho\otimes\sigma$ in $\mathcal{H}_{A}\otimes
\mathcal{H}_{B}$ such that $\Lambda$\ yields outcome $1$ with probability at
least $\eta>0$ when applied to $\rho\otimes\sigma$. \ Then if we apply
$\Lambda_{j_{1}},\ldots,\Lambda_{j_{T}}$\ in sequence to $\rho$, where
$j_{1},\ldots,j_{T}$ are drawn uniformly and independently from $\left\{
1,\ldots,N\right\}  $ and $T\geq N/\eta^{2}$, the probability that at least
one of these\ measurements yields outcome $1$ is at least $\left(  \eta
-\sqrt{N/T}\right)  ^{2}$.
\end{lemma}

\begin{proof}
Let $E_{t}$\ denote the event that one of the first $t$ measurements of $\rho
$\ yields outcome $1$. \ Also, let $\alpha:=\left(  \eta-\sqrt{N/T}\right)
^{2}$. \ Then our goal is to show that $\Pr\left[  E_{t}\right]  \geq\alpha$
for some $t$, where the probability is over the choice of $j_{1},\ldots,j_{T}$
as well as the measurement outcomes. \ Suppose $\Pr\left[  E_{t}\right]
<\alpha$ for all $t$; we will derive a contradiction.

Let $\rho_{t}$\ be the state in $\mathcal{H}_{A}$\ after the first $t$
measurements, averaged over all choices of $j_{1},\ldots,j_{t}$\ and assuming
$E_{t}$ does not occur. \ Suppose $\left\Vert \rho_{t}-\rho\right\Vert
_{\operatorname*{tr}}>\sqrt{\alpha}$\ for some $t$. \ Then interpreting the
first $t$ measurements as a single measurement, and taking the contrapositive
of Lemma \ref{goodasnew}, we find that $\Pr\left[  E_{t}\right]  >\alpha$, and
we are done. \ So we can assume without loss of generality that $\left\Vert
\rho_{t}-\rho\right\Vert _{\operatorname*{tr}}\leq\sqrt{\alpha}$\ for all $t$.

For all mixed states $\varsigma$ in $\mathcal{H}_{A}\otimes\mathcal{H}_{B}$,
let $P_{\Lambda}\left(  \varsigma\right)  $\ be the probability that $\Lambda
$\ yields outcome $1$ when applied to $\varsigma$. \ By the definition of
trace distance, we have%
\[
P_{\Lambda}\left(  \varsigma^{\prime}\right)  \geq P_{\Lambda}\left(
\varsigma\right)  -\left\Vert \varsigma-\varsigma^{\prime}\right\Vert
_{\operatorname*{tr}}%
\]
for all $\varsigma,\varsigma^{\prime}$. \ Therefore%
\begin{align*}
P_{\Lambda}\left(  \rho_{t}\otimes\sigma\right)   &  \geq P_{\Lambda}\left(
\rho\otimes\sigma\right)  -\left\Vert \rho_{t}\otimes\sigma-\rho\otimes
\sigma\right\Vert _{\operatorname*{tr}}\\
&  =P_{\Lambda}\left(  \rho\otimes\sigma\right)  -\left\Vert \rho_{t}%
-\rho\right\Vert _{\operatorname*{tr}}\\
&  \geq\eta-\sqrt{\alpha}.
\end{align*}
Hence%
\[
P_{\Lambda}\left(  \rho_{t}\otimes I\right)  \geq\frac{\eta-\sqrt{\alpha}}%
{N},
\]
where%
\[
I=\frac{1}{N}\sum_{j=1}^{N}\left\vert j\right\rangle \left\langle
j\right\vert
\]
is the maximally mixed state in $\mathcal{H}_{B}$. \ It follows that for all
$t$,%
\[
\operatorname*{EX}_{j\in\left\{  1,\ldots,N\right\}  }\left[  P_{\Lambda
}\left(  \rho_{t}\otimes\left\vert j\right\rangle \left\langle j\right\vert
\right)  \right]  \geq\frac{\eta-\sqrt{\alpha}}{N}.
\]
Now\ notice that%
\[
\Pr\left[  E_{t}|\urcorner E_{t-1}\right]  =\operatorname*{EX}_{j_{t}%
\in\left\{  1,\ldots,N\right\}  }\left[  P_{\Lambda}\left(  \rho_{t-1}%
\otimes\left\vert j_{t}\right\rangle \left\langle j_{t}\right\vert \right)
\right]
\]
for all $t$. \ Furthermore, since $E_{t-1}\Rightarrow E_{t}$, the events
$\urcorner E_{t-1}\wedge E_{t}$\ are disjoint. \ Therefore%
\begin{align*}
\Pr\left[  E_{T}\right]   &  =\sum_{t=1}^{T}\Pr\left[  \urcorner E_{t-1}\wedge
E_{t}\right] \\
&  =\sum_{t=1}^{T}\Pr\left[  \urcorner E_{t-1}\right]  \Pr\left[
E_{t}|\urcorner E_{t-1}\right] \\
&  \geq\sum_{t=1}^{T}\left(  1-\alpha\right)  \cdot\operatorname*{EX}%
_{j_{t}\in\left\{  1,\ldots,N\right\}  }\left[  P_{\Lambda}\left(  \rho
_{t-1}\otimes\left\vert j_{t}\right\rangle \left\langle j_{t}\right\vert
\right)  \right] \\
&  \geq\left(  1-\alpha\right)  T\left(  \frac{\eta-\sqrt{\alpha}}{N}\right)
\\
&  \geq\left(  \eta-\sqrt{\alpha}\right)  ^{2}\frac{T}{N}\\
&  =1,
\end{align*}
which is certainly greater than $\alpha=\left(  \eta-\sqrt{N/T}\right)  ^{2}$.
\ Here we are using the fact that $T\geq N/\eta^{2}$, and hence $\alpha\leq1$.
\end{proof}

\SubSection{Amplification\label{AMP}}

Before proceeding further, we need to decrease Bob's soundness error (that is,
the probability that he accepts a dishonest claim from Merlin). \ The simplest
approach would be to have Alice and Merlin both send $\ell$\ copies of their
messages for some $\ell$, and then have Bob run his verification algorithm
$\ell$\ times in parallel and output the majority answer. \ However, this
approach fails, since the decrease in error probability is more than cancelled
out by the \textit{increase} in Merlin's message length (recall that we will
have to loop over all possible classical messages from Merlin). \ So then why
not use the \textquotedblleft in-place amplification\textquotedblright%
\ technique of Marriott and Watrous \cite{mw}? \ Because unfortunately, that
technique only works for Merlin's message; we do not know whether it can be
generalized to handle Alice's message as well.\footnote{In any such
generalization, certainly Alice will still have to send multiple copies of her
message. \ The question is whether Merlin will also have to send multiple
copies of \textit{his} message.} \ Happily, there is a \textquotedblleft
custom\textquotedblright\ amplification procedure with the properties we want:

\begin{lemma}
\label{doubleamp}Suppose Bob receives an $a$-qubit message $\left\vert
\psi\right\rangle $ from Alice and a $w$-qubit message $\left\vert
\varphi\right\rangle $\ from Merlin, where $w\geq2$. \ Let $A=O\left(
aw\log^{2}w\right)  $\ and $W=O\left(  w\log w\right)  $. \ Then by using $A$
qubits from Alice and $W$ qubits from Merlin, Bob can amplify his soundness
error to $5^{-W}$ while keeping his completeness error $1/3$.
\end{lemma}

\begin{proof}
We will actually use two layers of amplification. \ In the \textquotedblleft
inner\textquotedblright\ layer, we replace Alice's message $\left\vert
\psi\right\rangle $ by the $a\ell$-qubit\ message $\left\vert \psi
\right\rangle ^{\otimes\ell}$, where $\ell=O\left(  \log w\right)  $. \ We
also replace Merlin's message $\left\vert \varphi\right\rangle $\ by the
$w\ell$-qubit message $\left\vert \varphi\right\rangle ^{\otimes\ell}$. \ We
then run Bob's algorithm $\ell$\ times in parallel and output the majority
answer. \ By a Chernoff bound, together with the same observations used by
Kitaev and Watrous \cite{kitaevwatrous}\ to show amplification for
$\mathsf{QMA}$, this reduces both the completeness and the soundness errors to
$\varepsilon=\frac{1}{1000w^{3}}$, for suitable $\ell=O\left(  \log w\right)
$.

In the \textquotedblleft outer\textquotedblright\ layer, we replace Alice's
message by $\left\vert \psi\right\rangle ^{\otimes\ell u}$, where $u=O\left(
W\right)  $. \ We then run the inner layer $u$\ times, once for each copy of
$\left\vert \psi\right\rangle ^{\otimes\ell}$, but reusing the same register
for Merlin's message\ each time. \ (Also, after each invocation of the inner
layer, we uncompute everything except the final answer.) \ Finally, we output
the majority answer among these $u$ invocations.

Call Bob's original algorithm $\mathcal{Q}$, and call the amplified algorithm
$\mathcal{Q}_{\ast}$. \ Then our first claim is that if $\mathcal{Q}$ accepts
all $w$-qubit\ messages from Merlin with probability at most $1/3$, then
$\mathcal{Q}_{\ast}$\ accepts all $W$-qubit messages\ with probability at most
$5^{-W}$, for suitable $u=O\left(  W\right)  $. \ This follows from a Chernoff
bound---since even if we condition on the first through $t^{th}$\ invocations
of the inner layer, the $\left(  t+1\right)  ^{st}$\ invocation will still
receive a \textquotedblleft fresh\textquotedblright\ copy of $\left\vert
\psi\right\rangle ^{\otimes\ell}$, and will therefore accept with probability
at most $\varepsilon\leq1/3$. \ The state of Merlin's message register before
the $\left(  t+1\right)  ^{st}$\ invocation is irrelevant.

Our second claim is that, if $\mathcal{Q}$ accepts some $\left\vert
\varphi\right\rangle $\ with probability at least $2/3$, then $\mathcal{Q}%
_{\ast}$\ accepts $\left\vert \varphi\right\rangle ^{\otimes\ell}$\ with
probability at least $2/3$. \ For recall that a single invocation of the inner
layer rejects $\left\vert \varphi\right\rangle ^{\otimes\ell}$ with
probability at most $\varepsilon$. \ So by Lemma \ref{unionbound}, even if we
invoke the inner layer $u$ times in sequence, the probability that one or more
invocations reject is at most $u\sqrt{\varepsilon}$, which is less than
$1/3$\ for suitable $u=O\left(  W\right)  $.
\end{proof}

\SubSection{Main Result\label{MAIN}}

We are now ready to prove Theorem \ref{demerlin}: that for all Boolean
functions $f$\ and all $w\geq2$,%
\[
\operatorname*{Q}\nolimits^{1}\left(  f\right)  =O\left(  \operatorname*{QMA}%
\nolimits_{w}^{1}\left(  f\right)  \cdot w\log^{2}w\right)  .
\]
Furthermore, if Bob uses $C$ gates and $S$\ qubits in the $\operatorname*{QMA}%
\nolimits_{w}^{1}$\ protocol, then he uses\ $C\cdot S^{O\left(  S\right)  }%
$\ gates and $O\left(  S^{2}\log^{2}S\right)  $\ qubits in the
$\operatorname*{Q}\nolimits^{1}$\ protocol.

\begin{proof}
[Proof of Theorem \ref{demerlin}]Let $\mathcal{Q}$\ be Bob's algorithm.
\ Also, suppose Alice's message has $a$ qubits and Merlin's message has $w$
qubits. \ The first step is to replace $\mathcal{Q}$ by the
amplified\ algorithm $\mathcal{Q}_{\ast}$\ from Lemma \ref{doubleamp}, which
takes an $A$-qubit advice state $\left\vert \Psi\right\rangle $\ from Alice
and a $W$-qubit witness state from Merlin, where $A=O\left(  aw\log
^{2}w\right)  $ and $W=O\left(  w\log w\right)  $. \ From now on, we use
$\mathcal{Q}_{\ast}\left(  \left\vert \Phi\right\rangle \right)  $\ as a
shorthand for $\mathcal{Q}_{\ast}$\ run with witness $\left\vert
\Phi\right\rangle $, together with an advice register that originally contains
Alice's message $\left\vert \Psi\right\rangle $ (but that might become
corrupted as Bob uses it). \ Then Bob's goal is to decide whether there exists
a $\left\vert \Phi\right\rangle $\ such that $\mathcal{Q}_{\ast}\left(
\left\vert \Phi\right\rangle \right)  $ accepts with high probability.

To do so, Bob uses the following procedure $\mathcal{M}$. \ Given Alice's
message $\left\vert \Psi\right\rangle $, this procedure runs $\mathcal{Q}%
_{\ast}\left(  \left\vert z\right\rangle \right)  $\ for $9\left(
2^{W}\right)  $ computational basis states $\left\vert z\right\rangle $ of the
witness register chosen uniformly at random. \ Finally it returns the logical
OR of the measurement outcomes.\bigskip

\qquad let $\left\vert c\right\rangle $\ be a counter initialized to
$\left\vert 0\right\rangle $

\qquad for $t:=1$\ to $9\left(  2^{W}\right)  $

\qquad\qquad choose $z\in\left\{  0,1\right\}  ^{W}$\ uniformly at random

\qquad\qquad run $\mathcal{Q}_{\ast}\left(  \left\vert z\right\rangle \right)
$, and let $b$\ be $\mathcal{Q}_{\ast}$'s output

\textit{\qquad\qquad\qquad// }$1$\textit{ for accept, }$0$\textit{ for reject}

\qquad\qquad set $\left\vert c\right\rangle :=\left\vert c+b\right\rangle $

\qquad\qquad run $\mathcal{Q}_{\ast}^{-1}\left(  \left\vert z\right\rangle
\right)  $\ to uncompute garbage

\qquad next $t$

\qquad if $c=0$\ then return $f\left(  x,y\right)  =0$;

\qquad\qquad otherwise return $f\left(  x,y\right)  =1\bigskip$

Let us first show that $\mathcal{M}$ is correct. \ First suppose that
$f\left(  x,y\right)  =0$. \ By Lemma \ref{doubleamp}, we know that
$\mathcal{Q}_{\ast}\left(  \left\vert \Phi\right\rangle \right)  $\ accepts
with probability at most $5^{-W}$\ for all states $\left\vert \Phi
\right\rangle $ of the witness register. \ So in particular, $\mathcal{Q}%
_{\ast}\left(  \left\vert z\right\rangle \right)  $\ accepts with probability
at most $5^{-W}$\ for all basis states $\left\vert z\right\rangle $. \ By
Lemma \ref{unionbound}, it follows that when $\mathcal{M}$\ is finished, the
counter $c$ will have been incremented at least once (and hence $\mathcal{M}%
$\ itself will have accepted) with probability at most%
\[
\frac{9\left(  2^{W}\right)  }{\sqrt{5^{W}}}\ll\frac{1}{9}.
\]
Next suppose that $f\left(  x,y\right)  =1$. \ By assumption, there exists a
$\left\vert \Phi\right\rangle $\ such that $\mathcal{Q}_{\ast}\left(
\left\vert \Phi\right\rangle \right)  $\ accepts with probability at least
$2/3$. \ So setting $\eta=2/3$, $N=2^{W}$, and $T=9\left(  2^{W}\right)  $,
Lemma \ref{orbound}\ implies that $\mathcal{M}$ will accept with probability
at least%
\[
\left(  \eta-\sqrt{\frac{N}{T}}\right)  ^{2}=\left(  \frac{2}{3}-\sqrt
{\frac{1}{9}}\right)  ^{2}=\frac{1}{9}.
\]

It remains only to upper-bound $\mathcal{M}$'s complexity. \ If Bob's original
algorithm $\mathcal{Q}$\ used $C$ gates and\ $S$ qubits, then clearly the
amplified algorithm $\mathcal{Q}_{\ast}$\ uses $O\left(  C\cdot w\log
^{2}w\right)  $\ gates and $O\left(  S\cdot w\log^{2}w\right)  $\ qubits.
\ Hence $\mathcal{M}$\ uses%
\[
O\left(  C\cdot w\log^{2}w\cdot2^{W}\right)  =C\cdot S^{O\left(  S\right)  }%
\]
gates and $O\left(  S^{2}\log^{2}S\right)  $\ qubits, where we have used the
fact that $w\leq S$. \ This completes the proof.
\end{proof}

\Section{Conclusions and Open Problems\label{OPEN}}%

%TCIMACRO{\FRAME{ftbpFU}{2.7605in}{2.5272in}{0pt}{\Qcb{Known containments among
%classical and quantum advice classes.}}{\Qlb{nonunif}}{nonunif.eps}%
%width 2.7605in;  height 2.5272in;  depth 0pt;  original-width 10.3511in;
%original-height 7.7551in;  cropleft "0.2673";  croptop "0.9473";
%cropright "0.7475";  cropbottom "0.3615";
%filename 'nonunif.eps';file-properties "XNPEU";}}}%
%BeginExpansion
\begin{figure}
[ptb]
\begin{center}
\includegraphics[
trim=2.766849in 2.803468in 2.613653in 0.408694in,
height=2.5272in,
width=2.7605in
]%
{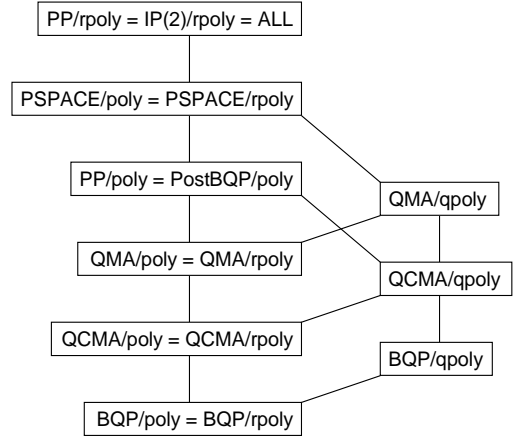}%
\caption{Known containments among classical and quantum advice classes.}%
\label{nonunif}%
\end{center}
\end{figure}
%EndExpansion
Figure \ref{nonunif} shows the known relationships among deterministic,
randomized, and quantum advice classes, in light of this paper's results. \ We
still know remarkably little about quantum advice, compared to other
computational resources. \ But our results provide new evidence for a general
hypothesis: that if you're strong enough to squeeze an exponential amount of
advice out of a quantum state, then you're also strong enough to squeeze an
exponential amount of advice out of a probability distribution.

We end with some open problems.

\begin{itemize}
\item Can we find a counterexample to the quantum advice hypothesis? \ What
about $\mathsf{QMA}\left(  2\right)  $, or $\mathsf{QMA}\left(  k\right)
$\ for $k>2$,\ or $\mathsf{QS}_{2}^{p}$? \ Currently, we do not even know
whether $\mathsf{QMA}\left(  2\right)  \mathsf{/rpoly}=\mathsf{ALL}$; this
seems related to the difficult open question of amplification for
$\mathsf{QMA}\left(  2\right)  $\ (see Kobayashi et al.\ \cite{kmy}).

\item Is there a class $\mathcal{C}$ such that $\mathcal{C}\mathsf{/rpoly}%
\neq\mathcal{C}\mathsf{/poly}$\ but $\mathcal{C}\mathsf{/rpoly}\neq
\mathsf{ALL}$?

\item Can we tighten the $\Omega\left(  N/\log^{2}N\right)  $\ lower bound
of\ Theorem \ref{qraclb}\ to $\Omega\left(  N\right)  $? \ One approach would
be to tighten Lemma \ref{doubleamp}, by generalizing the in-place
$\mathsf{QMA}$\ amplification of Marriott and Watrous \cite{mw}.

\item Can we improve the containment $\mathsf{QMA/qpoly}\subseteq
\mathsf{PSPACE/poly}$\ to $\mathsf{QMA/qpoly}\subseteq\mathsf{PP/poly}$?
\ Alternatively, can we construct an oracle (possibly a `quantum
oracle'\ \cite{ak}) relative to which $\mathsf{QMA/qpoly}\not \subset
\mathsf{PostBQP/poly}$? \ This would indicate that the upper bound of
$\mathsf{PSPACE/poly}$\ might be difficult to improve.
\end{itemize}

\Section{Acknowledgments}

Greg Kuperberg collaborated in the research project of which this paper was an
offshoot, and I am grateful to him for comments and advice, as well as for
several observations including Proposition \ref{maexp}. \ I also thank Oded
Regev for first suggesting to me the problem of proving an upper bound on
$\mathsf{QMA/qpoly}$; Ashwin Nayak and Hirotada Kobayashi for pointing out
errors in an earlier version of Section \ref{QIL}; Harumichi Nishimura for
helpful discussions; Jon Yard for pointing out a gap in an earlier version of
Section \ref{QMAPP}; and Ronald de Wolf and the anonymous reviewers for
comments on the manuscript.

\bibliographystyle{latex8}
\bibliography{thesis}

\begin{thebibliography}{10}\setlength{\itemsep}{-1ex}\small

\bibitem{aar:adv}
S.~Aaronson.
\newblock Limitations of quantum advice and one-way communication.
\newblock {\em Theory of Computing}, 1:1--28, 2005.
\newblock quant-ph/0402095.

\bibitem{aar:pp}
S.~Aaronson.
\newblock Quantum computing, postselection, and probabilistic polynomial-time.
\newblock {\em Proc. Roy. Soc. London}, A461(2063):3473--3482, 2005.
\newblock quant-ph/0412187.

\bibitem{ak}
S.~Aaronson and G.~Kuperberg.
\newblock Quantum versus classical proofs and advice.
\newblock In preparation, 2006.

\bibitem{antv}
A.~Ambainis, A.~Nayak, A.~Ta-Shma, and U.~V. Vazirani.
\newblock Quantum dense coding and quantum finite automata.
\newblock {\em J. ACM}, 49:496--511, 2002.
\newblock Earlier version in ACM STOC 1999, pp. 376-383. quant-ph/9804043.

\bibitem{gs}
S.~Goldwasser and M.~Sipser.
\newblock Private coins versus public coins in interactive proof systems.
\newblock In {\em Randomness and Computation}, volume~5 of {\em Advances in
  Computing Research}. JAI Press, 1989.

\bibitem{kitaevwatrous}
A.~Kitaev and J.~Watrous.
\newblock Parallelization, amplification, and exponential-time simulation of
  quantum interactive proof systems.
\newblock In {\em Proc. ACM STOC}, pages 608--617, 2000.

\bibitem{klauck:cc}
H.~Klauck.
\newblock Quantum communication complexity.
\newblock In {\em Proc. Intl. Colloquium on Automata, Languages, and
  Programming (ICALP)}, pages 241--252, 2000.
\newblock quant-ph/0005032.

\bibitem{kmy}
H.~Kobayashi, K.~Matsumoto, and T.~Yamakami.
\newblock Quantum {M}erlin-{A}rthur proof systems: are multiple {M}erlins more
  helpful to {A}rthur?
\newblock In {\em ISAAC}, pages 189--198, 2003.
\newblock quant-ph/0306051.

\bibitem{ladner:ppspace}
R.~E. Ladner.
\newblock Polynomial space counting problems.
\newblock {\em SIAM J. Comput.}, 18:1087--1097, 1989.

\bibitem{mw}
C.~Marriott and J.~Watrous.
\newblock Quantum {A}rthur-{M}erlin games.
\newblock {\em Computational Complexity}, 14(2):122--152, 2005.

\bibitem{nayak}
A.~Nayak.
\newblock Optimal lower bounds for quantum automata and random access codes.
\newblock In {\em Proc. IEEE FOCS}, pages 369--377, 1999.
\newblock quant-ph/9904093.

\bibitem{newman}
I.~Newman.
\newblock Private vs. common random bits in communication complexity.
\newblock {\em Inform. Proc. Lett.}, 39:67--71, 1991.

\bibitem{nc}
M.~Nielsen and I.~Chuang.
\newblock {\em Quantum Computation and Quantum Information}.
\newblock Cambridge University Press, 2000.

\bibitem{raz:all}
R.~Raz.
\newblock Quantum information and the {PCP} theorem.
\newblock In {\em Proc. IEEE FOCS}, 2005.
\newblock quant-ph/0504075.

\bibitem{watrous:space}
J.~Watrous.
\newblock Space-bounded quantum complexity.
\newblock {\em J. Comput. Sys. Sci.}, 59(2):281--326, 1999.

\end{thebibliography}

\Section{Appendix: Other Complexity Results\label{APPENDIX}}

The purpose of this appendix is to show that, in upper-bounding
$\mathsf{QMA/qpoly}$, the computational difficulty really does arise from the
need to handle quantum advice and quantum witnesses simultaneously:\ if either
or both are \textquotedblleft dequantized,\textquotedblright\ then the upper
bound of $\mathsf{PSPACE/poly}$\ can be improved. \ In particular, and in
increasing order of nontriviality, Theorem \ref{marlog}\ will show that
$\mathsf{MA/rpoly}=\mathsf{MA/poly}$ (and likewise that $\mathsf{QCMA/rpoly}%
=\mathsf{QCMA/poly}$), Theorem \ref{qmarpoly}\ will show that
$\mathsf{QMA/rpoly}=\mathsf{QMA/poly}$, and Theorem \ref{qcmaqpolythm}\ will
show that $\mathsf{QCMA/qpoly}\subseteq\mathsf{PP/poly}$.

First, however, let us make a cautionary observation, which illustrates why
such upper bounds cannot be blithely assumed. \ Recall that $\mathsf{MA}%
_{\mathsf{EXP}}$ is the exponential-time analogue of $\mathsf{MA}$.

\begin{proposition}
\label{maexp}$\mathsf{MA}_{\mathsf{EXP}}\mathsf{/rpoly}=\mathsf{ALL}$.
\end{proposition}

\begin{proof}
Given an arbitrary Boolean function $f:\left\{  0,1\right\}  ^{n}%
\rightarrow\left\{  0,1\right\}  $, an honest Merlin's message will consist of
the truth table of $f$, while the randomized advice will consist of an
$O\left(  n\right)  $-bit fingerprint of the truth table.
\end{proof}

We can also \textquotedblleft scale down\textquotedblright\ Proposition
\ref{maexp}\ by an exponential, to obtain $\mathsf{MA/poly}\subseteq
\mathsf{MA/rl{}og}$. \ More explicitly, in the $\mathsf{MA/rl{}og}%
$\ simulation, an honest Merlin's message will contain the advice $s$ to the
$\mathsf{MA/poly}$\ machine, while the $\mathsf{rl{}og}$ advice will consist
of an $O\left(  \log n\right)  $-bit fingerprint of $s$.

We next show that $\mathsf{MA/rpoly}=\mathsf{MA/poly}$. \ Combined with the
above observation, this result has the surprising implication that%
\[
\mathsf{MA/rl{}og}=\mathsf{NP/poly}=\mathsf{MA/poly}=\mathsf{MA/rpoly.}%
\]
In other words, for an $\mathsf{MA}$\ machine, $\operatorname*{poly}\left(
n\right)  $\ bits of randomized advice are no more powerful than $\log\left(
n\right)  $\ bits.

\begin{theorem}
\label{marlog}$\mathsf{MA/rpoly}=\mathsf{MA/poly}$.
\end{theorem}

\begin{proof}
Let $L$ be a language in $\mathsf{MA/rpoly}$, and let $\mathcal{A}\left(
x,r,z\right)  $\ be Arthur's verification algorithm run on input $x$, advice
string $r$, and witness $z\in\left\{  0,1\right\}  ^{w\left(  n\right)  }$,
for some polynomial $w$. \ (We assume without loss of generality that Arthur
is deterministic, since the randomized\ advice can provide his coins.) \ Also,
let $\mathcal{D}$\ be the distribution from which $r$ is drawn. \ Then for all
$x\in L$, there exists a $z$ such that%
\[
\Pr_{r\in\mathcal{D}}\left[  \mathcal{A}\left(  x,r,z\right)  \text{
accepts}\right]  \geq\frac{2}{3},
\]
whereas for all $x\notin L$\ and all $z$,%
\[
\Pr_{r\in\mathcal{D}}\left[  \mathcal{A}\left(  x,r,z\right)  \text{
accepts}\right]  \leq\frac{1}{3}.
\]
Let $R=\left(  r_{1},\ldots,r_{p\left(  n\right)  }\right)  $\ be a $p\left(
n\right)  $-tuple of independent samples\ from $\mathcal{D}$, for some
$p\left(  n\right)  =\Theta\left(  n+w\left(  n\right)  \right)  $. \ Then
there exists a boosted verifier $\mathcal{A}_{\ast}$ such that for all $x\in
L$, there exists a $z$ such that%
\[
\Pr_{R\in\mathcal{D}^{p\left(  n\right)  }}\left[  \mathcal{A}_{\ast}\left(
x,R,z\right)  \text{ accepts}\right]  \geq1-\frac{1}{2^{n}2^{w\left(
n\right)  }},
\]
whereas for all $x\notin L$\ and all $z$,%
\[
\Pr_{R\in\mathcal{D}^{p\left(  n\right)  }}\left[  \mathcal{A}_{\ast}\left(
x,R,z\right)  \text{ accepts}\right]  \leq\frac{1}{2^{n}2^{w\left(  n\right)
}}.
\]
So by a simple counting argument, there exists a fixed advice string $R$\ such
that for all $x\in L$, there exists a $z$ such that Arthur accepts; whereas
for all $x\notin L$\ and all $z$, Arthur rejects.
\end{proof}

Indeed, using the same techniques we can show that%
\[
\mathsf{QCMA/rl{}og}=\mathsf{QCMA/ql{}og}=\mathsf{QCMA/poly}%
=\mathsf{QCMA/rpoly}.
\]

Next we want to show a somewhat harder result, that $\mathsf{QMA/rpoly}%
=\mathsf{QMA/poly}$. \ To do so we will need the following theorem of Marriott
and Watrous.

\begin{theorem}
[Marriott and Watrous \cite{mw}]\label{mwthm}The error probability in any
$\mathsf{QMA}$\ protocol can be made exponentially small without increasing
the size of Merlin's quantum witness.
\end{theorem}

We can now prove the analogue of Theorem \ref{marlog} for $\mathsf{QMA}$.

\begin{theorem}
\label{qmarpoly}$\mathsf{QMA/rpoly}=\mathsf{QMA/poly}.$
\end{theorem}

\begin{proof}
Given a language $L\in\mathsf{QMA/rpoly}$, let $\mathcal{D}$\ be the
distribution from which Arthur's\ advice is drawn, and let $\mathcal{Q}\left(
x,r,\left\vert \varphi\right\rangle \right)  $\ be Arthur's verification
algorithm run on input $x$, advice string $r$, and witness $\left\vert
\varphi\right\rangle \in\mathcal{H}_{2}^{\otimes w\left(  n\right)  }$. \ Then
for all $x\in L$, there exists a $\left\vert \varphi\right\rangle $ such that%
\[
\Pr_{r\in\mathcal{D}}\left[  \mathcal{Q}\left(  x,r,\left\vert \varphi
\right\rangle \right)  \text{ accepts}\right]  \geq\frac{2}{3},
\]
whereas for all $x\notin L$\ and all $\left\vert \varphi\right\rangle $,%
\[
\Pr_{r\in\mathcal{D}}\left[  \mathcal{Q}\left(  x,r,\left\vert \varphi
\right\rangle \right)  \text{ accepts}\right]  \leq\frac{1}{3}.
\]
Here the probability is taken over $\mathcal{Q}$'s internal randomness as well
as $r$.

By Theorem \ref{mwthm}, we can make the error probability exponentially small
without increasing the size of $\left\vert \psi\right\rangle $. \ So let
$R=\left(  r_{1},\ldots,r_{p\left(  n\right)  }\right)  $\ be a $p\left(
n\right)  $-tuple of independent samples\ from $\mathcal{D}$, for some
$p\left(  n\right)  =\Theta\left(  n+w\left(  n\right)  \right)  $. \ Then
there exists a boosted verifier $\mathcal{Q}_{\ast}$ such that for all $x\in
L$, there exists a $\left\vert \varphi\right\rangle $ such that%
\[
\Pr_{R\in\mathcal{D}^{p\left(  n\right)  }}\left[  \mathcal{Q}_{\ast}\left(
x,R,\left\vert \varphi\right\rangle \right)  \text{ accepts}\right]
\geq1-\frac{1}{2^{n}2^{3w\left(  n\right)  }},
\]
whereas for all $x\notin L$\ and all $\left\vert \varphi\right\rangle $,%
\[
\Pr_{R\in\mathcal{D}^{p\left(  n\right)  }}\left[  \mathcal{Q}_{\ast}\left(
x,R,\left\vert \varphi\right\rangle \right)  \text{ accepts}\right]  \leq
\frac{1}{2^{n}2^{3w\left(  n\right)  }}.
\]
So by a simple counting argument, there exists a fixed advice string $R_{1}%
$\ such that for all $x\in L$, there exists a $\left\vert \varphi\right\rangle
$ such that Arthur accepts with probability at least $1-2^{-3w\left(
n\right)  }$. \ However, we still need to handle the\ case $x\notin L$.
\ Since the number of states $\left\vert \varphi\right\rangle \in
\mathcal{H}_{2}^{\otimes w\left(  n\right)  }$\ with small pairwise inner
product is \textit{doubly} exponential, a na\"{\i}ve counting argument no
longer works. \ Instead, observe that there exists a fixed advice string
$R_{0}$\ such that for all $x\notin L$\ and all computational basis states
$\left\vert z\right\rangle $ with $z\in\left\{  0,1\right\}  ^{w\left(
n\right)  }$,%
\begin{align*}
\Pr\left[  \mathcal{Q}_{\ast}\left(  x,R_{0},\left\vert z\right\rangle
\right)  \text{ accepts}\right]   &  \leq2^{n}2^{w\left(  n\right)  }%
\cdot\frac{1}{2^{n}2^{3w\left(  n\right)  }}\\
&  =\frac{1}{2^{2w\left(  n\right)  }}.
\end{align*}
Now suppose by contradiction that there exists a $\left\vert \varphi
\right\rangle $\ such that%
\[
\Pr\left[  \mathcal{Q}_{\ast}\left(  x,R_{0},\left\vert \varphi\right\rangle
\right)  \text{ accepts}\right]  >\frac{1}{3}.
\]
Then%
\[
\Pr\left[  \mathcal{Q}_{\ast}\left(  x,R_{0},I\right)  \text{ accepts}\right]
>\frac{1}{3}\cdot\frac{1}{2^{w\left(  n\right)  }},
\]
where%
\[
I=\frac{1}{2^{w\left(  n\right)  }}\sum_{z\in\left\{  0,1\right\}  ^{w\left(
n\right)  }}\left\vert z\right\rangle \left\langle z\right\vert
\]
is the maximally mixed state on $w\left(  n\right)  $\ qubits. \ But this
implies that there exists a basis state $\left\vert z\right\rangle $\ such
that%
\[
\Pr\left[  \mathcal{Q}_{\ast}\left(  x,R_{0},\left\vert z\right\rangle
\right)  \text{ accepts}\right]  >\frac{1}{3}\cdot\frac{1}{2^{w\left(
n\right)  }},
\]
which yields the desired contradiction. \ Finally, by a union bound, there
exists a fixed advice string $R$\ that combines the properties of $R_{0}$\ and
$R_{1}$.
\end{proof}

\SubSection{\label{QCMASEC}Upper-Bounding QCMA/qpoly}

We now show that $\mathsf{QCMA/qpoly}\subseteq\mathsf{PP/poly}$.
\ Conceptually, the proof is similar to the proof that $\mathsf{QMA/qpoly}%
\subseteq\mathsf{PSPACE/poly}$, but with three differences. \ First, since the
witnesses are now classical, they can be provided to the simulating machine as
part of the advice. \ Second, since the witnesses are provided, there is no
longer any need to try exponentially many random witnesses. \ Indeed, this is
what improves the upper bound from $\mathsf{PSPACE/poly}$\ to
$\mathsf{PP/poly}$. \ And third, we can no longer exploit the fact that
$\mathsf{BQPSPACE/qpoly}=\mathsf{PSPACE/poly}$, in order to split the proof
neatly into a \textquotedblleft de-Merlinization\textquotedblright\ part
(which is new) and an \textquotedblleft advice\textquotedblright\ part (which
follows from earlier work of Aaronson \cite{aar:adv}). \ Instead, we need to
generalize the machinery from \cite{aar:adv} to the $\mathsf{QCMA}$ setting.

\begin{theorem}
\label{qcmaqpolythm}$\mathsf{QCMA/qpoly}\subseteq\mathsf{PP/poly}$.
\end{theorem}

\begin{proof}
Let $L$ be a language in $\mathsf{QCMA/qpoly}$, and let $L\left(  x\right)
=1$\ if $x\in L$\ and $L\left(  x\right)  =0$\ otherwise. \ Also, let
$\mathcal{Q}$ be a verifier for $L$, which takes a $a$-qubit quantum
advice\ state $\left\vert \psi\right\rangle $\ and $w$-bit classical witness
$z$\ for some polynomials $a$ and $w$ (for convenience, we omit the dependence
on $n$). \ Then the first step is to replace $\mathcal{Q}$\ by an amplified
verifier $\mathcal{Q}_{\ast}$, which takes an $A$-qubit advice state
$\left\vert \Psi\right\rangle :=\left\vert \psi\right\rangle ^{\otimes\ell}$,
where $A=a\ell$ and $\ell=O\left(  \log a\right)  $. \ As a result,
$\mathcal{Q}_{\ast}$\ has completeness and soundness errors $1/A^{4}$.

Let $\mathcal{Q}_{\ast}\left(  x,\rho,z\right)  $\ be shorthand for
$\mathcal{Q}_{\ast}$\ run with input $x$, advice $\rho$, and witness $z$.
\ Then given $x$, our goal is to simulate $\mathcal{Q}_{\ast}\left(
x,\left\vert \Psi\right\rangle ,z\left(  x\right)  \right)  $, where $z\left(
x\right)  $\ is an optimal witness for $x$. \ We will do so\ using a
$\mathsf{PP/poly}$ machine $\mathcal{M}$. \ The classical advice to
$\mathcal{M}$ will consist of a \textquotedblleft Darwinian training
set\textquotedblright\ $\left(  x_{1},z_{1}\right)  ,\ldots,\left(
x_{T},z_{T}\right)  $\ for $T=O\left(  A\right)  $, together with $L\left(
x_{t}\right)  $ for every $t\in\left\{  1,\ldots,T\right\}  $. \ Here each
$x_{t}\in\left\{  0,1\right\}  ^{n}$\ is an input and each $z_{t}\in\left\{
0,1\right\}  ^{w}$\ is its corresponding witness. \ Given this advice,
$\mathcal{M}$ runs the following procedure to compute $L\left(  x\right)
$.\bigskip

let $\rho:=I_{A}$ be the maximally mixed state on $A$\ qubits

for $t:=1$\ to $T$

\qquad let $\left\vert b\right\rangle $\ be a qubit initialized to $\left\vert
0\right\rangle $

\qquad run $\mathcal{Q}_{\ast}\left(  x_{t},\rho,z_{t}\right)  $, and CNOT the
output into $\left\vert b\right\rangle $

\qquad run $\mathcal{Q}_{\ast}^{-1}\left(  x_{t},\rho,z_{t}\right)  $ to
uncompute garbage

\qquad measure $\left\vert b\right\rangle $, and postselect on observing
$b=L\left(  x_{t}\right)  $

next $t$

for all $z\in\left\{  0,1\right\}  ^{w}$,

\qquad let $\lambda_{z}$\ be the probability that $\mathcal{Q}_{\ast}\left(
x,\rho,z\right)  $\ accepts

if there exists a $z$\ such that $\lambda_{z}\geq2/3$, then accept

otherwise, if $\lambda_{z}\leq1/3$\ for all $z$, then reject\bigskip

Let us first see why $\mathcal{M}$\ can be simulated in $\mathsf{PP/poly}$.
\ The `for' loop is just a postselected quantum computation, and can clearly
be simulated by the result of Aaronson \cite{aar:pp}\ that $\mathsf{PostBQP}%
=\mathsf{PP}$. \ The one nontrivial step is to decide whether there exists a
$z$\ such that $\lambda_{z}\geq2/3$, or whether $\lambda_{z}\leq1/3$\ for all
$z$. \ We do this as follows. \ Let $\rho_{t}$\ be the state of the advice
register after the first $t$ postselection steps, conditioned on those steps
succeeding. \ We first amplify by repeating the `for' loop $J=O\left(
w\right)  $\ times, using a different advice register\ each time. \ This
yields $J$ copies of $\rho_{T}$. \ We then replace $\mathcal{Q}_{\ast}\left(
x,\rho_{T},z\right)  $\ by the doubly-amplified verifier $\mathcal{Q}_{\ast
}^{\prime}\left(  x,\rho_{T}^{\otimes J},z\right)  $, which runs
$\mathcal{Q}_{\ast}\left(  x,\rho_{T},z\right)  $\ once for each of the $J$
advice registers, and returns the majority outcome. \ Let $\lambda_{z}%
^{\prime}$\ be the probability that $\mathcal{Q}_{\ast}^{\prime}\left(
x,\rho_{T}^{\otimes J},z\right)  $\ accepts. \ Then by a Chernoff bound, and
assuming the constant in $J=O\left(  w\right)  $\ is sufficiently large, we
have reduced the problem to deciding whether

\begin{enumerate}
\item[(1)] there exists a $z\in\left\{  0,1\right\}  ^{w}$ such that
$\lambda_{z}^{\prime}\geq1-2^{-2w}$, or

\item[(2)] $\lambda_{z}^{\prime}\leq2^{-2w}$\ for all $z$.
\end{enumerate}

Now let%
\[
S:=\frac{1}{2^{w}}\sum_{z\in\left\{  0,1\right\}  ^{w}}\lambda_{z}^{\prime}.
\]
Then $S\geq2^{-w-1}$ in case (1), whereas $S\leq2^{-2w}$ in case (2). \ So it
suffices to give a $\mathsf{PP/poly}$\ machine with $\alpha+\beta
S$\ accepting paths, for some positive constants $\alpha$\ and $\beta$. \ Our
machine will simply do the following:

\begin{itemize}
\item Choose $z$\ uniformly at random.

\item Simulate a $\mathsf{PostBQP}$\ computation that accepts with probability
proportional to $\lambda_{z}^{\prime}$.
\end{itemize}

The reason this works is that the probability of the $T$ postselection steps
in the `for' loop all succeeding is independent of $z$.

It remains only to show $\mathcal{M}$'s correctness. \ Let $p_{t}$\ be the
probability that the first $t$ postselection steps in the `for' loop all
succeed. \ We choose the \textquotedblleft training inputs\textquotedblright%
\ $x_{1},\ldots,x_{T}$\ and witnesses $z_{1},\ldots,z_{T}$\ in such a way that

\begin{enumerate}
\item[(a)] $p_{t+1}\leq\frac{2}{3}p_{t}$\ for all $t\in\left\{  0,\ldots
,T-1\right\}  $.

\item[(b)] $z_{t}$\ is a valid witness for $x_{t}$ whenever $x_{t}\in L$,
meaning that $\mathcal{Q}_{\ast}\left(  x_{t},\left\vert \Psi\right\rangle
,z_{t}\right)  $\ accepts with probability at least $1-1/A^{4}$.

\item[(c)] There is no larger training set that satisfies (a) and (b).
\end{enumerate}

Then it suffices to prove the following two claims:

\begin{enumerate}
\item[(i)] $T=O\left(  A\right)  $\ for all training sets that satisfy (a) and (b).

\item[(ii)] $\mathcal{M}$\ correctly decides every input $x$, if we train it
on some $\left(  x_{1},z_{1}\right)  ,\ldots,\left(  x_{T},z_{T}\right)
$\ that satisfies (a), (b), and (c).
\end{enumerate}

For Claim (i), notice that we can write the maximally mixed state $I$\ as a
mixture of $2^{A}$\ orthonormal vectors%
\[
I=\frac{1}{2^{A}}\sum_{i=1}^{2^{A}}\left\vert \Psi_{i}\right\rangle
\left\langle \Psi_{i}\right\vert ,
\]
where $\left\vert \Psi_{1}\right\rangle :=\left\vert \Psi\right\rangle $\ is
the \textquotedblleft true\textquotedblright\ advice state. \ We argue that
the $\left\vert \Psi_{1}\right\rangle \left\langle \Psi_{1}\right\vert
$\ component must survive all $T$ postselection steps with high probability.
\ For if $x_{t}\notin L$, then $\mathcal{Q}_{\ast}\left(  x_{t},\left\vert
\Psi\right\rangle ,z_{t}\right)  $\ accepts with probability at most $1/A^{4}%
$, while if $x_{t}\in L$, then $\mathcal{Q}_{\ast}\left(  x_{t},\left\vert
\Psi\right\rangle ,z_{t}\right)  $\ rejects with probability at most $1/A^{4}$
by assumption (b). \ So by Lemma \ref{unionbound}, the probability of
outputting the wrong answer on any of $\left(  x_{1},z_{1}\right)
,\ldots,\left(  x_{T},z_{T}\right)  $, using $\left\vert \Psi\right\rangle
$\ as the advice, is at most $T\sqrt{1/A^{4}}=T/A^{2}$. \ Hence%
\[
p_{T}\geq\frac{1}{2^{A}}\left(  1-\frac{T}{A^{2}}\right)  .
\]
On the other hand, $p_{t+1}\leq\frac{2}{3}p_{t}$\ for all $t$ by assumption
(a), and hence $p_{T}\leq\left(  2/3\right)  ^{T}$. \ Combining we obtain
$T=O\left(  A\right)  $.

For Claim (ii), suppose by way of contradiction that $\mathcal{M}$\ rejects
some $x\in L$. \ Then $\mathcal{Q}_{\ast}\left(  x,\rho_{T},z\right)
$\ accepts with probability less than $2/3$ for all $z$. \ But this implies
that if we trained $\mathcal{M}$\ on the enlarged set $\left(  x_{1}%
,z_{1}\right)  ,\ldots,\left(  x_{T},z_{T}\right)  ,\left(  x,z\right)  $ for
any $z$, then we would get $p_{T+1}\leq\frac{2}{3}p_{T}$, thereby
contradicting the maximality of $T$. \ Likewise, suppose $\mathcal{M}%
$\ accepts some $x\notin L$. \ Then there exists a \textquotedblleft false
witness\textquotedblright\ $\widehat{z}$\ such that $\mathcal{Q}_{\ast}\left(
x,\rho_{T},\widehat{z}\right)  $\ accepts with probability greater than $1/3$.
\ So if we trained $\mathcal{M}$\ on the enlarged set $\left(  x_{1}%
,z_{1}\right)  ,\ldots,\left(  x_{T},z_{T}\right)  ,\left(  x,\widehat
{z}\right)  $, we would again get $p_{T+1}\leq\frac{2}{3}p_{T}$, contradicting
the maximality of $T$.
\end{proof}

\end{document}